\documentclass[10pt,twocolumn]{article}

\usepackage{units}
\usepackage{subcaption}
\usepackage{footmisc}
\usepackage{graphicx}%
\usepackage{multirow}%
\usepackage{amsmath,amssymb,amsfonts}%
\usepackage{amsthm}%
\usepackage{mathrsfs}%
\usepackage[figuresright]{rotating}%
\usepackage[title]{appendix}%
\usepackage{xcolor}%
\usepackage{textcomp}%
\usepackage{booktabs}%
\usepackage{algorithm}%
\usepackage{algorithmicx}%
\usepackage{algpseudocode}%
\usepackage{program}%
\usepackage{listings}%
\usepackage{authblk}
\usepackage[section]{placeins}
\usepackage{float}
\usepackage{hyperref}

\hypersetup{
	colorlinks=true,
	allcolors=blue,
}

\begin{document}

\title{\textsc{SurfFlow}: high-throughput surface energy calculations for arbitrary crystals}

\providecommand{\keywords}[1]
{
	\small	
	\textbf{\textit{Keywords---}} #1
}

\author[1]{Firat Yalcin}
\author[2]{Michael Wolloch}
\affil[1]{Computational Materials Physics, University of Vienna, Kolingasse 14-16, 1090, Vienna, Austria}
\affil[2]{Vasp Software GmbH, Sensengasse 8/12, 1090, Vienna, Austria}

\maketitle

\begin{abstract}
	We introduce \textsc{SurfFlow}, an open-source high-throughput workflow package designed for automated first-principles calculations of surface energies in arbitrary crystals. Our package offers a comprehensive solution capable of handling multi-element crystals, nonstoichiometric compositions, and asymmetric slabs, for all potential terminations. To streamline the computational process, \textsc{SurfFlow} employs an efficient pre-screening method that discards surfaces with suspected high surface energy before conducting resource-intensive density functional theory computations. The results generated are seamlessly compiled into an \textsc{optimade}-compliant database, ensuring easy access and compatibility. Additionally, a user-friendly web interface facilitates workflow submission and management, provides result visualization, and enables the examination of Wulff shapes. \textsc{SurfFlow} represents a valuable tool for researchers looking to explore surface energies and their implications in a diverse range of systems.
\end{abstract}


\section{Introduction}
\label{sec:int}
The surface energy of a material determines its stability and influences its adhesion properties, catalytic ability, and ability to form thin films and interfaces. It is defined as the work needed to create a new surface from a bulk crystal. It is a fundamental property that can be used to explain the stability of surface facets, the Wulff shape~\cite{Wulff1901}, which is the equilibrium crystal shape of the material, and phenomena such as surface reconstruction~\cite{Somorjai1994, Oura2013Surface}, segregation~\cite{Wynblatt1977, Seah1979, Polak2000}, and catalysis~\cite{Hammer2000, Stamenkovic2006, Norskov2009}. Catalytic properties are closely tied to surface energy, and nanomaterials with high surface energies show exceptional properties for electrocatalysis, photocatalysis, and gas sensor applications~\cite{Zhou2011}. The surface energies of a pair of materials are also an important descriptor for the adhesion energy of these materials, a property of great importance for the construction of solid-state batteries~\cite{Seymour2023, Restuccia2023}.

Experimental data on surface energies are scarce because of the technical difficulty of the measurements. The available data are primarily limited to specific facets of elemental crystals. Experimental measurements of surface energies of solids can be based e.g.~on cleavage~\cite{Obreimoff1930, Gilman1960}, interfacial energy of small elastic smooth spheres~\cite{Kendall1987}, or contact angle measurements of various liquids~\cite{Kwok1999, Kozbial2014}. The main problem with cleavage-based methods is plastic deformation at the crack tip, while the second approach strictly works only for amorphous solids. Contact angle measurements, on the other hand, are associated with several difficulties and limitations that affect the accuracy of results, such as surface contamination, roughness, and assumptions in the underlying mathematical models.

In the quest to find new and highly functional materials, expensive and time-consuming experimental studies have been supplemented more and more by large-scale computational high-throughput (HT) screenings in recent years, mainly employing density functional theory (DFT). Due to the difficulties with conducting experiments to determine surface energies, and their crucial importance for a range of applications (e.g.~Refs.~\cite{Williams2006, Xiao2020}), a couple of HT tools to tackle this problem have been published in recent years~\cite{Tran2016, Yang2020, Brlec2021, Palizhati2019}. Calculating surface properties via ab-initio atomistic modeling alleviates the problem of isolating specific facets and provides exceptional control of the system parameters.

Tran et al.~generated a database of the surface energies of elemental crystals~\cite{Tran2016}, in which they provide surface energies of more than 100 polymorphs of approximately 70 elements. Yang et al.~provide an open-source code~\cite{Yang2020} to generate organic surfaces from bulk molecular crystals, which was very recently updated to better interface with DFT and neural network interatomic potentials, among other improvements~\cite{Moayedpour2023}. Brlec et al.~provide a python library~\cite{Brlec2021} that automates the surface cleaving and the processing of raw DFT outputs to extract materials properties such as surface energy. Furthermore, there has been an interest in predicting surface properties from unrelaxed surfaces, skipping DFT relaxations, and using a neural network model to learn and predict cleavage energies and Wulff constructions~\cite{Palizhati2019}.

Although these studies are effective in calculating surface energies for specific systems, they are hampered by some constraints: the ability to handle multi-element crystals is very limited, and/or they are not fully automated, requiring the user to manually initiate calculations and run post-processing scripts. This hinders the high-throughput screening approach as manual handling of such a large number of systems quickly becomes impractical.

The main problem that so far eluded a comprehensive HT treatment of surface energies of multi-element crystals is to correctly and automatically handle asymmetric and/or nonstoichiometric slabs, which are unavoidable for most systems, while still keeping the computational effort tractable. This is further complicated by the sheer amount of symmetrically in-equivalent surface directions (Miller indices), which might have several unique terminations each. Looking at surfaces with Miller indices of only up to a maximal index of 3, a rather simple crystal with only two distinct elements might have on the order of 100 unique surface configurations, while only a small fraction of those will show favorable surface energy and actually contribute to the Wulff shape.

In the present work, we aim to solve these issues and present a fully automatic approach to calculate surface energies and Wulff shapes for arbitrary crystals. Our code package efficiently handles multiple terminations of asymmetric and nonstoichiometric slabs. Additionally, it can filter out polar surfaces\footnote{Polar surfaces usually have higher surface energies compared to non-polar ones and usually do not contribute to the Wulff shape. The surface energy of polar surfaces can be calculated with \textsc{SurfFlow} if wanted, however, as discussed in section S3 of the SM.}, and predict low energy ones, as well as provide fully automatic calculation handling, error correction, database operations, and output visualization.

The package is developed in Python 3 and builds on well-known and commonly used tools for material science and HT computations like \textit{pymatgen}~\cite{ONG2013314}, \textit{atomate}~\cite{MATHEW2017140}, and \textit{Fireworks}~\cite{jain:15}. It is available from the Python package index (PyPI) or a GitLab repository as open source software and uses the Vienna Ab-initio Simulation Package, \textsc{vasp}~\cite{Kresse1993, Kresse1996, Kresse1996a}, to perform the DFT calculations. See Sec.~\ref{sec:dft_settings} for the settings used. While the methods and algorithms used in our workflow are not individually new, we believe that no other freely available Python package offers a comparable richness of features, ease of use, and efficiency.

In the following sections, we will present our methodology, highlight our measures to optimize slab size and pre-screen for low-energy surfaces using bond valence information, and benchmark some results against available experimental and calculated data. A detailed flowchart and explanation of the workflow architecture can be found in S5 of the SM.

\section{Results \& Discussion}\label{sec:results_and_discussion}

\subsection{High-throughput handling of arbitrary surfaces}
To model a surface with a DFT code, which generally employs periodic boundary conditions in all directions, requires the construction of a slab, a thin slice of material separated by a vacuum region from its duplicates in one direction. The simplest case of a surface energy ($\gamma$) calculation is for a symmetric and stoichiometric slab, meaning that both surfaces of the slab are equivalent and the system contains an integer number of formula units of the primitive bulk cell of the material. Then the surface energy $\gamma$ is,

\begin{equation}
	\gamma = \frac{1}{2A}\left[E_\mathrm{slab} - N E_{bulk}\right] \quad,
\end{equation}

where $A$ is the area of the exposed surface, $E_\mathrm{slab}$ the total energy of the relaxed slab, $E_{bulk}$ the energy of the bulk unit\footnote{It has been shown that using a bulk cell with the same lateral symmetry helps to converge the surface energy quickly with respect to slab thickness~\cite{SUN201353}.}, and $N$ the number of formula units in the slab. For all other cases of symmetry and stoichiometry, a more generalized expression can be given

\begin{equation}
	\gamma_1 = \frac{1}{A}\left[E_\mathrm{slab} - \sum_i N_i\mu_i(p, T)\right] - \gamma_2 \quad,
	\label{eqn:gen_surfen}
\end{equation}

where asymmetry leads to two distinct surface energies given by $\gamma_1$ and $\gamma_2$, and the bulk energy is replaced by a term involving chemical potentials $\mu_i$, which quantifies the contributions of the missing atoms in nonstoichiometric slabs. Because the chemical potential of a species is a function of pressure and temperature, instead of a single value for the surface energy, we can talk about a region of stability for the surface that is defined by the environment.

While it is in principle possible to perform reference calculations to determine the region of stability with respect to chemical potentials, this gets progressively more difficult as the number of species in the system increases. It is also not always trivial to choose a bulk or gas-phase reference~\cite{Jia2019, Jia2021}.
We have thus decided to compute our surface energies with respect to a complete vacuum and not make an attempt to include chemical potentials.

There have been a number of ways presented in the literature to computationally decouple surfaces in asymmetric cases, without relying on chemical potentials. Notable examples are the wedge method~\cite{Zhang2016, Ma2019}, energy density method~\cite{Chetty1992}, surface passivation method~\cite{Zhang2016, Kaminski2017}, twinned slab method~\cite{Bruno2021}, and methods based on combinations of unrelaxed and relaxed surface energies~\cite{Heifets2001, Tian2018}. All these approaches intend to isolate one side of the slab by eliminating the contribution to the free energy of the other side.

In this work, we have chosen to employ the method by Tian et al., which calculates the surface energy as the combination of cleavage (very similar to the unrelaxed surface energy given in ~\cite{Heifets2001}) and relaxation energies in a way that is highly transferable and able to deal with both asymmetric and nonstoichiometric slabs~\cite{Tian2018}. This approach is also similar to that of Eglitis and Vanderbilt~\cite{Eglitis2007} with some improvements in dealing with asymmetric surfaces.
The approach is based on the notion that a slab is first cleaved from a bulk, and then the surfaces relax into their final shape.
The cleavage energy is defined as

\begin{equation}\label{eqn:cleavage}
	E_\mathrm{cleavage} = \frac{1}{2A} \left[E_\mathrm{slab}^\mathrm{unrelax} - N E_\mathrm{bulk}\right] \quad,
\end{equation}
where $E_\mathrm{slab}^\mathrm{unrelax}$ is the total energy of the unrelaxed slab, $E_\mathrm{bulk}$ is the energy of the bulk reference structure, and $N$ is the number of formula units in the slab. (We should note that this definition of the cleavage energy differs from its more common interpretation as the average surface energy of an asymmetric slab.) Relaxation energy for symmetric slabs is simply given as

\begin{equation}\label{eqn:relaxation}
	E_\mathrm{relaxation} = \frac{1}{2A}\left[E_\mathrm{slab}^\mathrm{relax} - E_\mathrm{slab}^\mathrm{unrelax} \right]\quad,
\end{equation}
where $E_\mathrm{slab}^\mathrm{relax}$ is the total energy of the fully relaxed slab.

Two key assumptions in this method allow it to treat asymmetric and nonstoichiometric slabs. First, it is assumed that the cleavage energy (sometimes referred to as the unrelaxed surface energy in literature) is equal for complementary terminations (i.e. sequential layers in the infinite bulk). This assumption follows from the idea that complementary terminations are created simultaneously by cleaving the bulk in a single plane\footnote{In reference~\cite{Tian2018}, this is claimed to hold only for systems where the constituent species have similar electronegativities. Based on some theoretical arguments and test calculations we believe that this is only true for extreme differences in electronegativity, but can be considered negligible for most systems. See the SM, section 4, for details.}.
Thus, one can calculate the contribution of the cleavage energy to the surface energy on each side of the slab separately, provided that the slab is stoichiometric, as Eqn.~\ref{eqn:cleavage} is valid only for stoichiometric slabs.

The second assumption is that freezing half of the slab is enough to decouple two surfaces of the slab and simulate a semi-infinite slab which is fulfilled if the slab is thick enough that the middle layers are bulk-like. The relaxation energy for asymmetric slabs can then be expressed separately for different terminations $T$

\begin{equation}\label{eqn:fixhalf}
	E_\mathrm{relaxation}(T) = \frac{1}{A}\left[E_\mathrm{slab}^\mathrm{relax}(T) - E_\mathrm{slab}^\mathrm{unrelax}\right] \quad.
\end{equation}

where this time we divide by $A$ since only one side of the slab is relaxed.

These assumptions together allow \textsc{SurfFlow} to compute the surface energy $\gamma$ as

\begin{equation}\label{eqn:surface_energy_surfgen}
    \gamma = E_\mathrm{cleavage} + E_\mathrm{relaxation} \quad ,
\end{equation}

where $E_\mathrm{relaxation}$ is either equation~\ref{eqn:relaxation} or equation~\ref{eqn:fixhalf}, depending on the symmetry of the slab. Of course, equation~\ref{eqn:surface_energy_surfgen} is dependent on the termination $T$ in the case of asymmetric slabs.

While it is possible in theory to handle all cases of symmetry and stoichiometry using this approach, in practice we have to impose further constraints on the systems to be able to automate the process of calculating surface energies.
First, our computational framework deals strictly with bulk-terminated surfaces, which means that adatoms or surface vacancies cannot be considered.

\begin{figure*}[h!]
	\centering
	\includegraphics [width=0.9\linewidth]{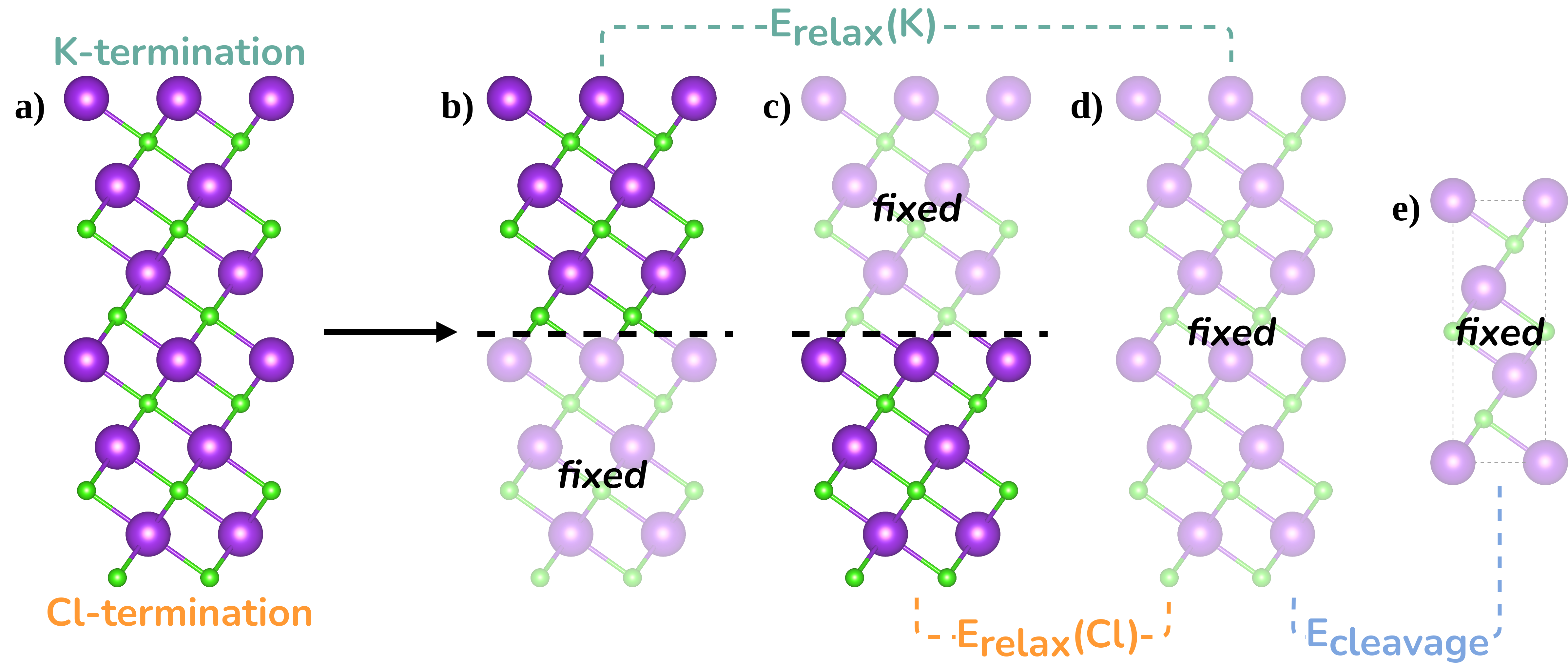}
	\caption{Surface energy calculation scheme for an asymmetric and stoichiometric KCl (mp-23193) (111) slab (a) with complementary terminations using the method by Tian et al.~\cite{Tian2018}. Calculations required are relaxations of the initial slab with only the (b) top, and (c) bottom halves relaxed, (c) static initial slab, and (d) static oriented unit cell.}
	\label{fig:asym_sto}
\end{figure*}

Furthermore, we limit the nonstoichiometric systems to symmetric slabs. This is done by modifying the bottom surfaces of slabs but is not possible for all slabs, thus this option is turned off by default and we generate equivalent stoichiometric, but asymmetric, slabs instead. This is possible in all cases and all possible terminations of a bulk are covered by this approach.
For these systems, it suffices to perform two relaxation calculations (top and bottom halves) and two static calculations (static slab and bulk reference) to calculate both surface energies $\gamma(T_1)$ and $\gamma(T_2)$. This case is depicted for an example system (KCl) in Fig.~\ref{fig:asym_sto}. For a more detailed explanation of the symmetric but nonstoichiometric case, see the SM, section~2.

With this method~\cite{Tian2018}, we are able to deal with all terminations using only a few calculations, avoid large system sizes (as for the wedge and twinned-slab methods), or face difficulties in generalizability (as for energy density and passivation methods).

\subsection{Optimizing performance by slab resizing and predicting low energy surfaces}

While simple DFT calculations with a GGA or metaGGA functional can no longer be considered expensive for systems up to around 100 atoms, slabs have usually low symmetry and especially higher index surfaces often need slabs with large lateral extensions. Even if a single calculation is relatively cheap, usually there are many unique directions with several possible terminations each, even for relatively simple crystals. Only surfaces with comparatively low energy are formed in most experimental circumstances, so those are of considerably more interest compared to unstable ones. It is therefore prudent to minimize the computational load by optimizing slabs and pre-screening surfaces to determine if it is likely that they might occur in experiments and/or contribute to the Wulff shape.

Optimization of slab size is a simpler task but still not trivial. While pymatgen provides great tools for generating slabs and minimizing their lateral dimensions, the minimal thickness is set by the size of the oriented unit cell (OUC). For high-index surfaces, this OUC might become quite large due to the necessity for periodic boundary conditions. For the slabs, the periodic boundary condition in the direction of the surface normal is broken by the vacuum however, \textsc{SurfFlow} reduces the thickness to match the intended layer count as accurately as possible. However, the original terminations must be preserved, which \textsc{SurfFlow} can achieve by clustering atomic sites into layers and removing appropriate chunks of them. This procedure is described in detail in S5.2 of the supplemental material (SM).

Filtering out surfaces that are unlikely to contribute to the Wulff shape because they should have very high surface energy is a much more complex task. In the first step, we attempt to filter out polar surfaces which usually have higher surface energies than their nonpolar counterparts of the same crystal~\cite{Noguera2000, Goniakowski2008, Dreyer2014}. The polarity of the surfaces is identified by pymatgen by guessing the most likely oxidation states and calculating the dipole moment in the direction of the surface normal. More about polar slabs can be found in section 3 of the SM.

Next, \textsc{SurfFlow} attempts to rank the remaining surfaces by quantifying the bonds broken during slab generation before actually performing any DFT calculations. Then, the workflow proceeds with $N$ candidates predicted to have low surface energies. These will be used for the Wulff shape generation. (Note that high-index and high energy surfaces can be beneficial for catalysis applications~\cite{Xiao2020}, so this option can be turned off when submitting a workflow to include all possible slabs compatible with the slab generation parameters.) In the following paragraphs, we will present the broken bond model used by \textsc{SurfFlow} in some more detail and benchmark the performance on a comprehensive set of materials with different structures, constituents, and bonding types.

Broken bond models have been used to predict surface energies of certain materials in the past~\cite{Methfessel1992, Galanakis2002, Gao2014}. However, the performance of the approach depends heavily on the material and the types of bonds present, which leads to better performance for some systems than for others. \textsc{SurfFlow} adapts this method to apply to a wide range of materials by counting and weighing broken bonds. This is especially important for multi-element crystals where bond strength can vary significantly.
The weights assigned to each bond should correspond to its strength, as the energy needed to break them is the main contributor to the surface energy.

\textsc{SurfFlow} makes use of bond valences that have been shown to approximate the bond energy by Etxebarria et al.~\cite{Etxebarria2005}. The valence of a bond between species $i$ and $j$ is given as

\begin{equation}
    S_{ij} = exp((R_0 - R_{ij})/b) \quad,
    \label{eqn:bond_valence}
\end{equation}

where $R_0$ is the optimal (for these bonding partners) and $R_{ij}$ is the realized bond length in the investigated material. The parameter $b$ measures the softness of the bond.

Eqn.~\ref{eqn:bond_valence} is the expression most widely used for bond valence, and tables of values for $R_{0}$ and $b$ for numerous bonds are available in the literature.
To be independent of necessarily incomplete tabulated data, \textsc{SurfFlow} uses the sum of the covalent radii of species forming the bond as the ideal bond length, and the frequently used \unit[0.37]{\AA} as $b$ parameter~\cite{Brown2009}\footnote{Some studies suggest that significantly different $b$ values should be used depending on bond type~\cite{Brown2009}, but this is intractable for an HT approach aimed at generality. However, we allow the default parameter to be changed in the setting file.}. We define the total bond valence sum (BVS) of a given site $i$ as a sum over all neighbors,

\begin{equation}
    S_{i} = \sum_{j \neq i} exp((R_0 - R_{ij})/b) \quad.
\end{equation}

We consider all neighbors up to 1.2 times the largest bond length of the bulk, however, due to the exponential decay of the bond valence, mostly nearest neighbors contribute.
The total bond valence sum of broken bonds $S$ is then simply a sum of all the partial sums over all surface sites,

\begin{equation}
    S = \sum_{i} S_{i} \quad.
\end{equation}

A similar, but slightly more complex approach was recently used to estimate the relative surface energies of homoatomic transition metal crystals with good success~\cite{Ma2020}. However, for our more diverse data set, the method presented here was found to be slightly more reliable.

No method based on bond breaking can differentiate between in-equivalent but complementary terminations of an asymmetric slab, because both surfaces originate from a single cleavage.
However, the bond valence sum correlates generally well with DFT surface energies, with a median Pearson correlation value of 0.87. Thus, in most cases, the method can be used as an excellent preliminary filter to weed out facets and terminations predicted to have high surface energies.

\subsection{Benchmarking the bond-valence model for predicting low energy surfaces}

\begin{figure}[h!]
	\centering
	\includegraphics[width=\linewidth]{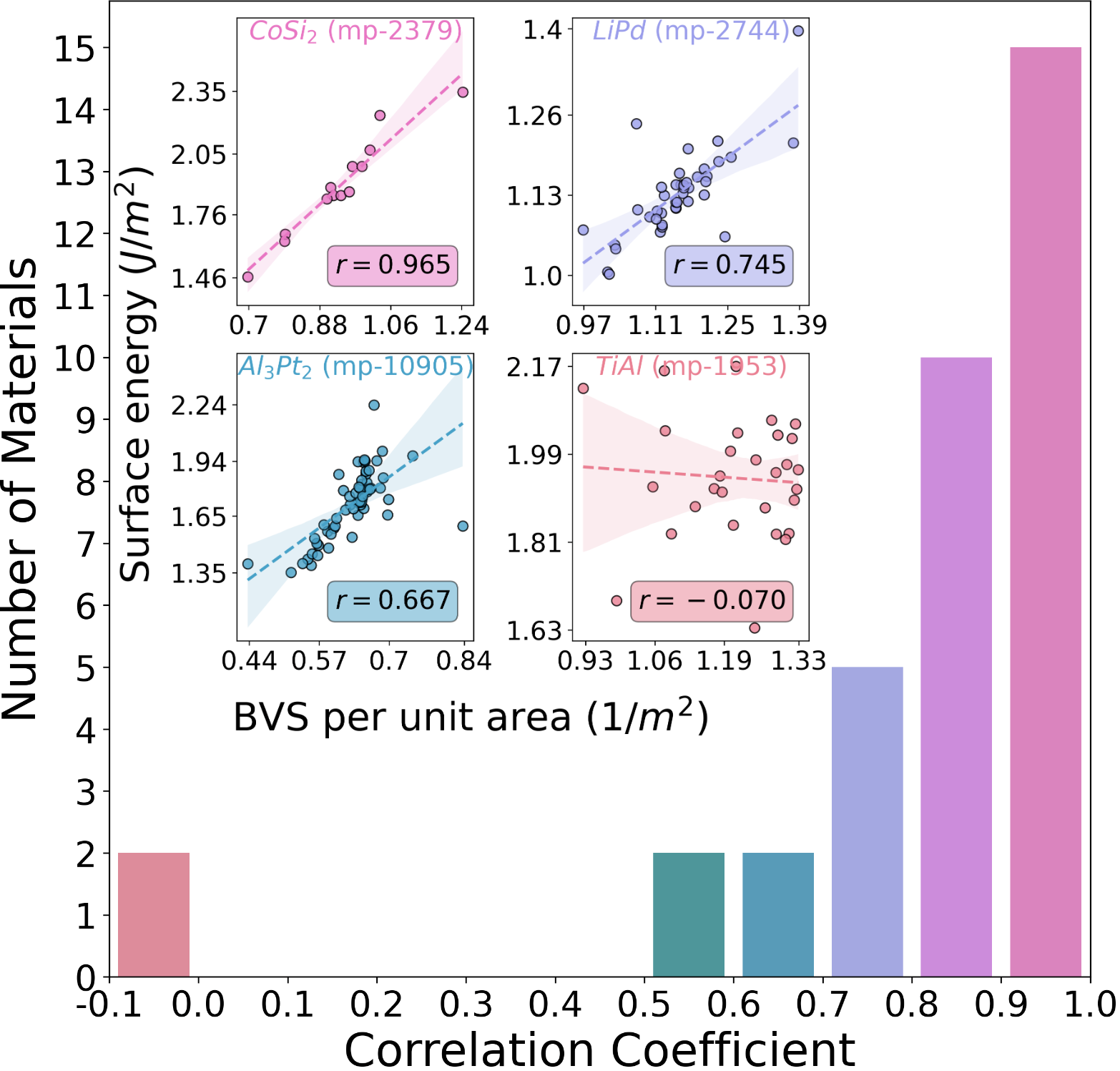}
	\caption{Main plot: Histogram sorting the investigated materials into bins according to the Pearson correlation numbers of the bond valence sum with the DFT surface energies. Inset: Surface energy $\gamma$ vs bond valence sum $S$ for four exemplary materials representing different success situations of the BVS estimation of the surface energy. Colors indicate the bins of the main plot. The shaded regions represent the 95\% confidence interval for the regression estimate.}
	\label{fig:bvs_vs_surfen}
\end{figure}

We have tested our bond valence model for 36 materials (See table~1 in the SM), encompassing different bonding characteristics, crystal structures, and constituting elements, from monoatomic to three-component systems. All unique non-polar surfaces up to a maximal Miller index (MMI) of 3 have been considered for each material, 653 surface energies in total\footnote{For a couple of systems we found only 4 non-polar unique surfaces, and have decided to include polar ones with $\mathrm{MMI} \leq 2$ as well for those. Also, slabs with more than 100 sites were disregarded to conserve computational resources. See S1 in the SM for more details.}. The median correlation coefficient across all test systems is $r_\mathrm{med}=0.870$, while two systems with slightly negative correlation reduce the average to $r_\mathrm{avg}=0.803$. However, 30 of the 36 materials tested show a correlation larger than 0.7, as can be seen in the histogram in Fig.~\ref{fig:bvs_vs_surfen}. The four worst performing systems are bcc Li (mp-135, $r=-0.087$), $\gamma$-TiAl (mp-1953, $r=-0.070$), as well as bcc and hcp Fe (mp-13, $r=0.573$; mp-136, $r=0.550$).
The by far worst correlation values (essentially 0) are observed for bcc Li and $\gamma$-TiAl. Both materials have several surfaces that are extremely close to one another in energy. For Li, a very soft metal, where all surface energies are low, and also the BVS values are very close together, this is not very surprising.
We thus recommend not relying on BVS filtering when calculating surface energies of very weakly bound materials.
For TiAl, however, the situation is different and might be rooted in difficulties in evaluating Ti $d-d$ interactions correctly in intermetallic systems. Sang and coworkers measured the charge density distribution of $\gamma$-TiAl with highly accurate quantitative convergent beam electron diffraction~\cite{Sang:2012}. While the qualitative agreement with full potential DFT calculations was good, there were considerable deviations for PAW-based calculations as we conduct them here.
A more detailed discussion about all systems with $p<0.6$ can be found in S11 of the SM.

In the inset of Fig.~\ref{fig:bvs_vs_surfen} we plot examples of the correlation between $S$ and $\gamma$ for four systems, colored according to the bins into which they are sorted.

To test the bond valence sum filtering approach on Wulff shapes, we use the $N$ surfaces with the lowest bond valence sums to construct Wulff shapes. All of the surface energies are calculated with DFT, however. We then compare the mean absolute errors (MAE) in the area fractions of these Wulff shapes with respect to the full DFT results up to the MMI 3.
The area fractions describe which percentage of the Wulff shape each facet contributes. 

\begin{figure}[H]
	\centering
	\begin{subfigure}{0.8\linewidth}
		\centering
		\includegraphics[width=\linewidth]{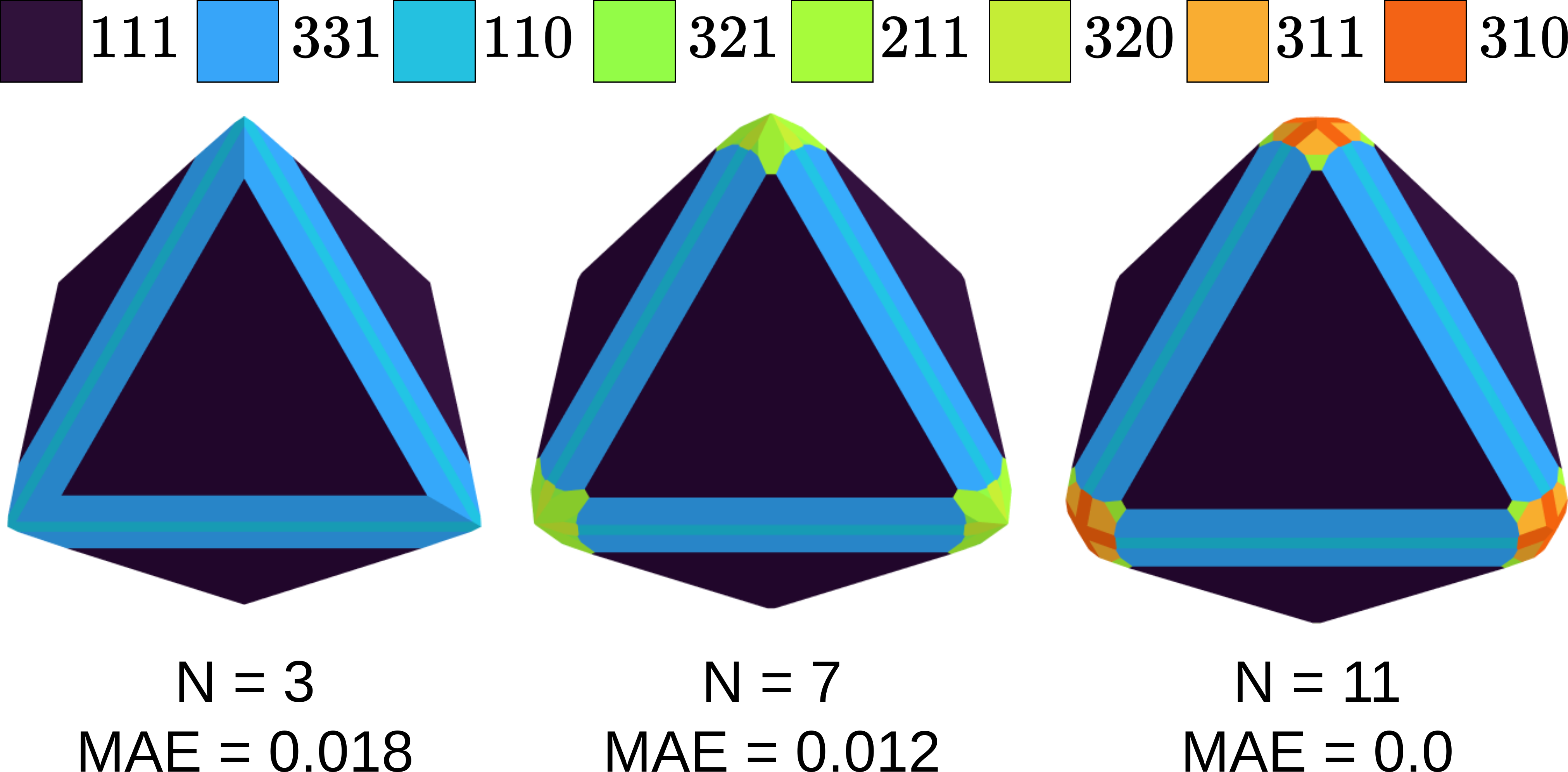}
		\caption{CoSi$_2$ (mp-2379)}
		\vspace{3mm}
		\label{fig:CoSi2_Wulff}
	\end{subfigure}
	\begin{subfigure}{0.8\linewidth}
		\centering
		\includegraphics[width=\linewidth]{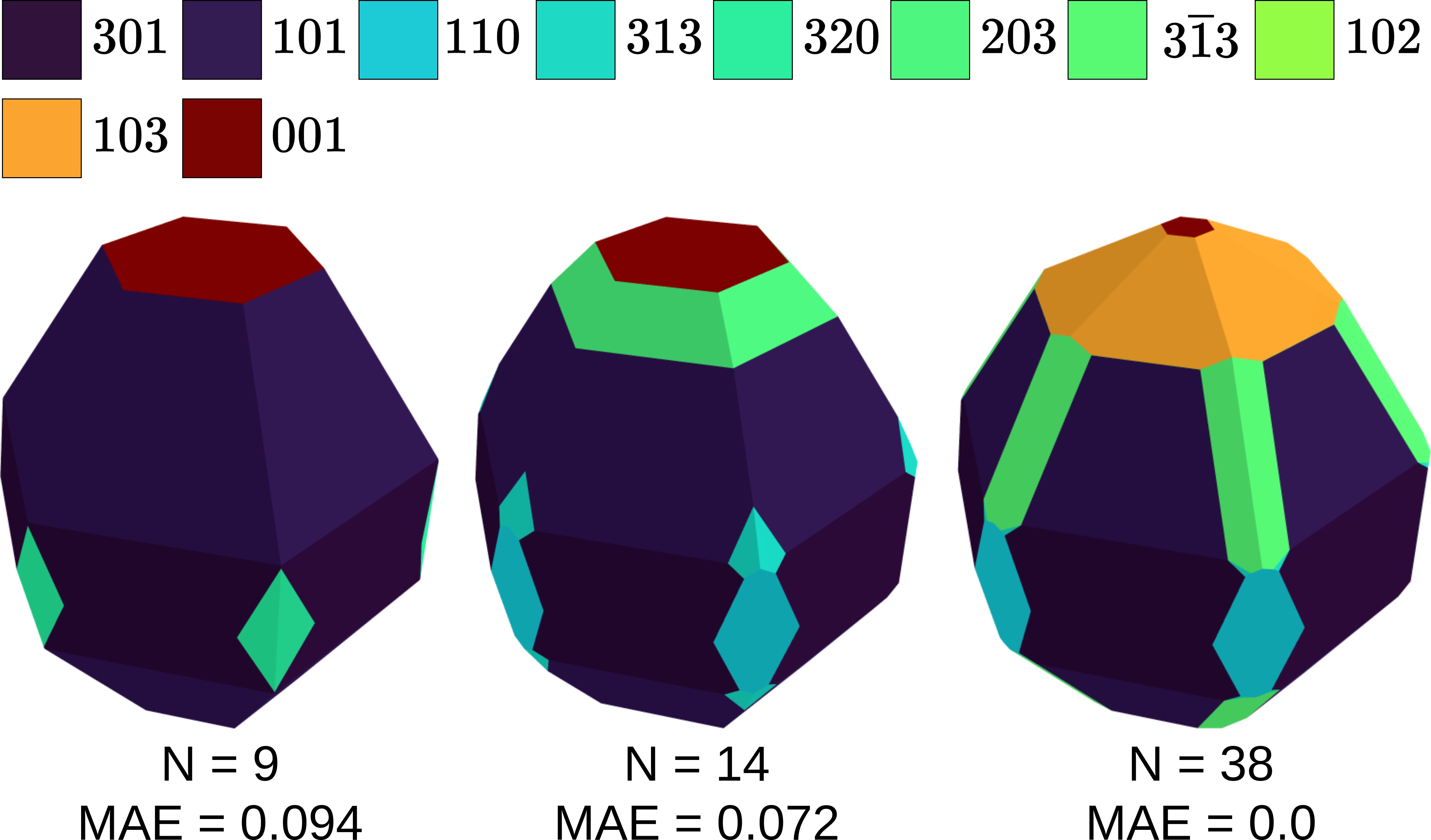}
		\caption{LiPd (mp-2744)}
		\vspace{3mm}
		\label{fig:LiPd_Wulff}
	\end{subfigure}
	
	\begin{subfigure}{0.8\linewidth}
		\centering
		\includegraphics[width=\linewidth]{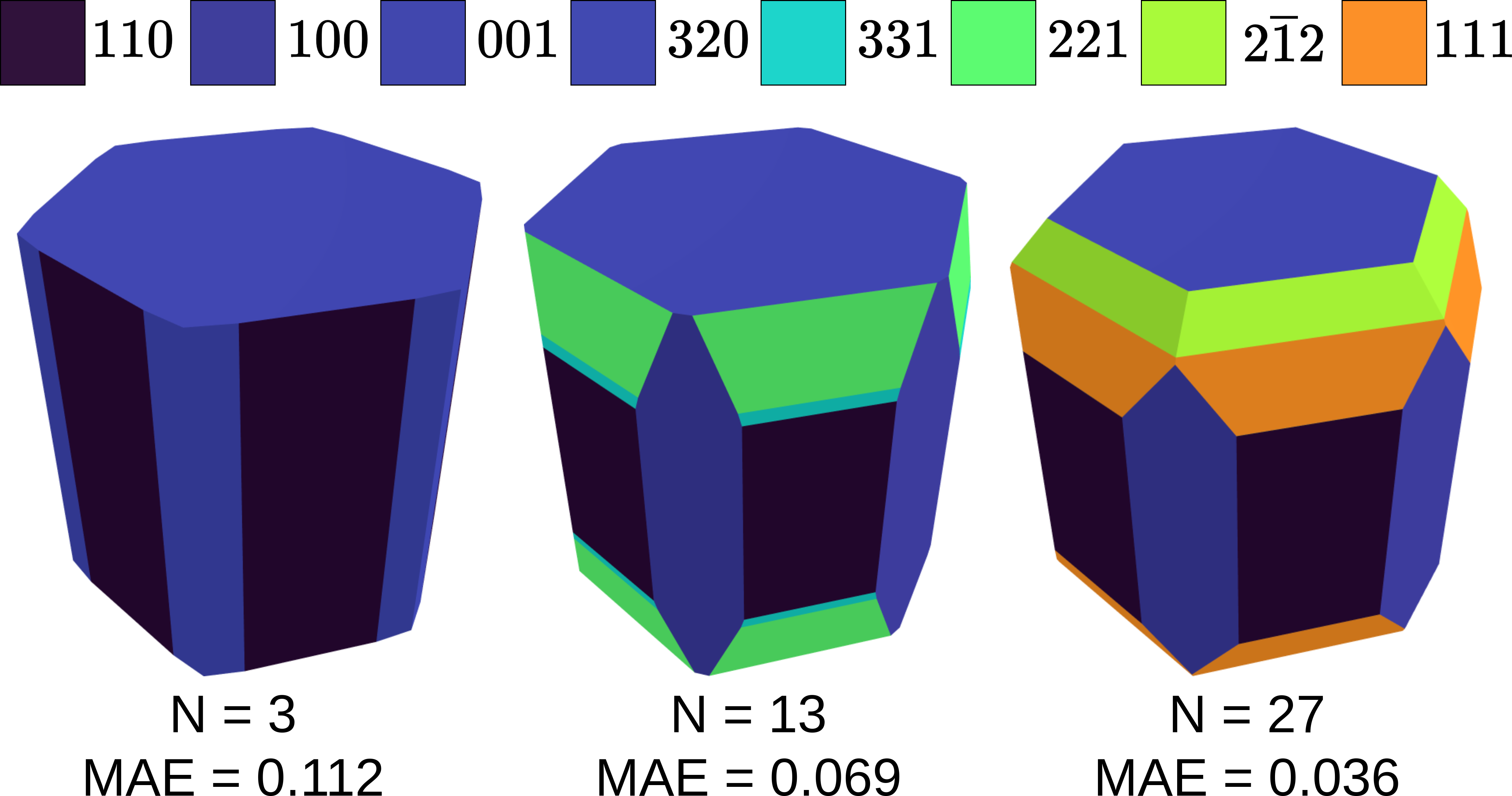}
		\caption{Al$_3$Pt$_2$ (mp-10905)}
		\vspace{3mm}
		\label{fig:Al2Pt3_Wulff}
	\end{subfigure}
	
	\begin{subfigure}{0.8\linewidth}
		\centering
		\includegraphics[width=\linewidth]{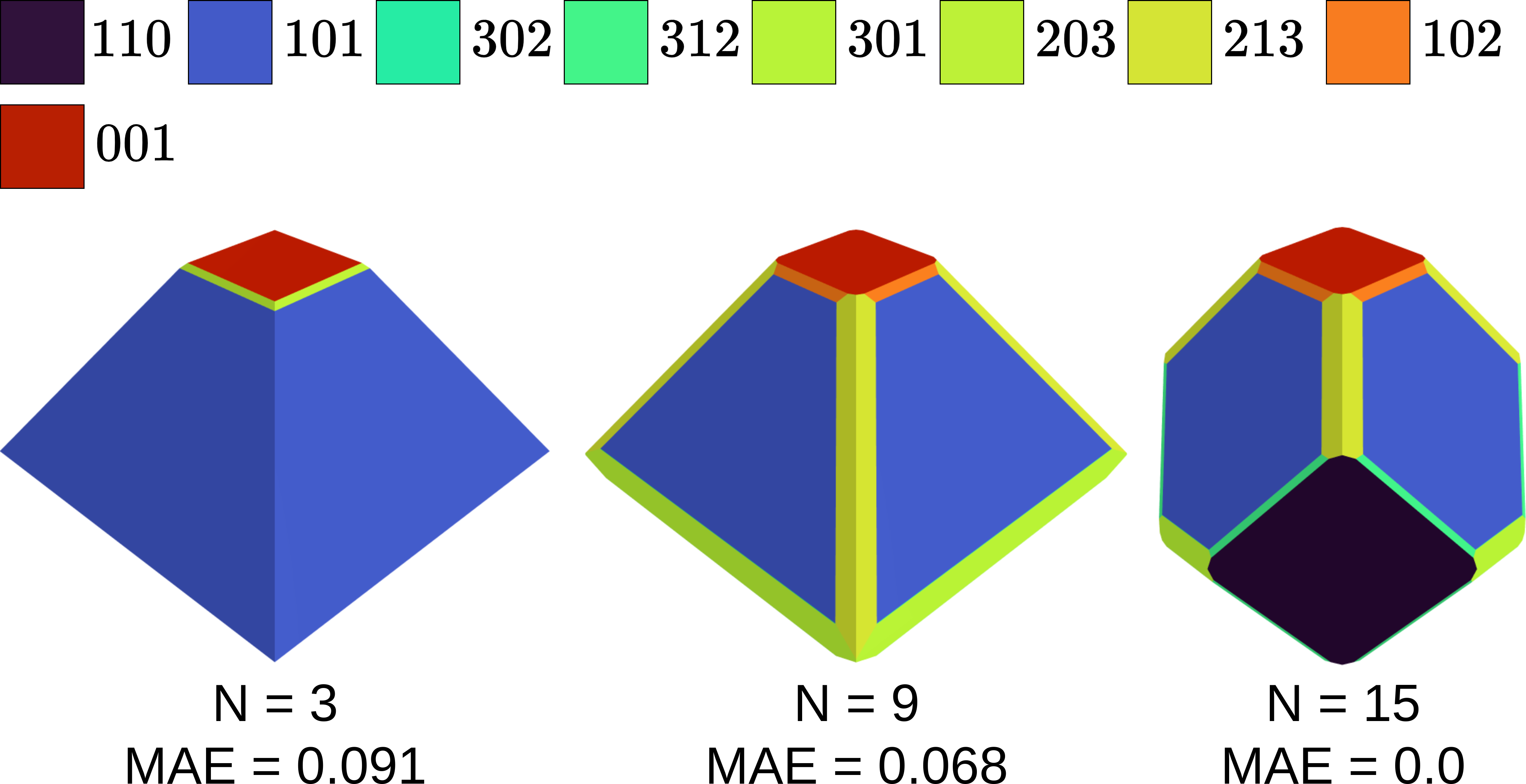}
		\vspace{1mm}
		\caption{TiAl (mp-1953)}
		\label{fig:TiAl_Wulff}
	\end{subfigure}
	\caption{Wulff shapes of 4 materials with varying correlations between $S$ and $\gamma$ for different numbers of calculated Miller-index/surface shift combinations. The facet colors correspond to the surface energy, from dark blue (lowest) to dark red (highest).}
	\label{fig:Wulff}
\end{figure}

In Fig.~\ref{fig:Wulff} we plot Wulff shapes for the four systems already presented in Fig.~\ref{fig:bvs_vs_surfen} calculated by DFT from the lowest $N$ BVS surfaces. These give examples of Wulff shape evolution if more and more surfaces are considered for different classes of BVS/$\gamma$ correlation.

In the case of CoSi$_2$ (Fig.~\ref{fig:CoSi2_Wulff}), where the correlation is really good, we see that for 3 computed surfaces, the (331) facet is quite prominently featured alongside the dominant (111) facet and a sliver of (110). The total MAE of the area fractions is already very low at 1.8\%. Adding four more surface energy calculations reduces the MAE a bit more by including (211), (320), and (321) facets, where the latter two disappear in the final shape in favor of the (311) and (310) facets that are both considerably higher in energy at $N=11$.

\begin{figure}[h!]
	\centering
	\includegraphics[width=\linewidth]{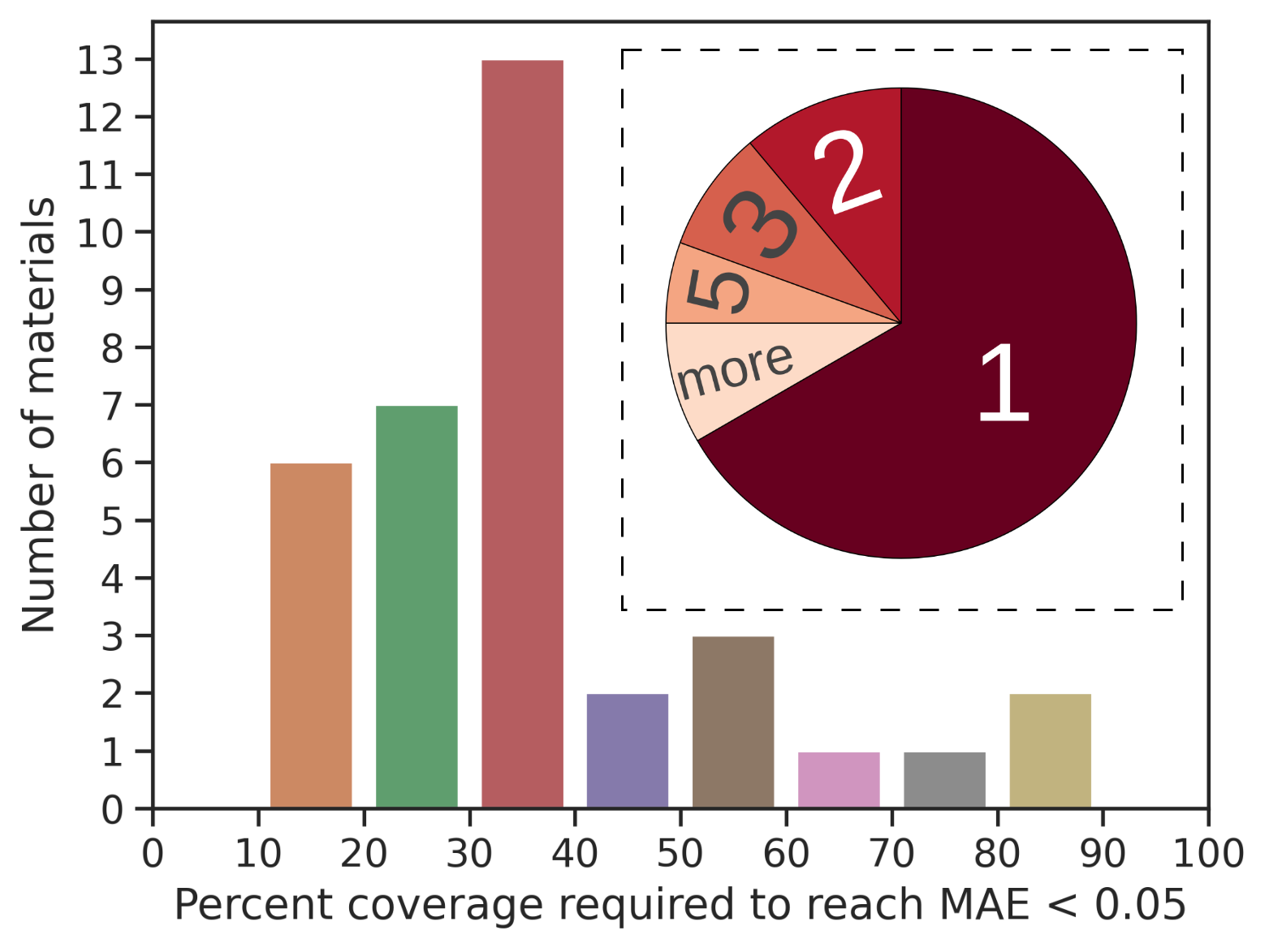}
	\caption{Coverage required to reach MAE threshold of 0.05 for area fraction compared to the full Wulff shape (coverage 100\%). The inset visualizes how many $N$ lowest BVS predictions must be calculated with DFT for all materials to find the lowest energy surface.}\label{fig:mae_barplot_wrt_coverage}
\end{figure}

For LiPd (Fig.~\ref{fig:LiPd_Wulff}) we have a total of 41 surfaces to consider. Again, a high index surface appears early with the (301) facet, which shows up even at $N=3$ and is ever so slightly lower in energy than (101). At $N=9$, the (320) and (001) facets also appear, but (320) ends up not contributing to the final shape at $N=38$, being replaced by the (101) surface. The (103) facet shown with considerable area fraction at $N=38$ does not appear until quite late at $N=32$, due to its relatively high surface energy. We see that with many surfaces close in energy, even systems with decent correlations between BVS and $\gamma$ might not produce the correct Wulff shape unless many surfaces are considered. MAEs are therefore higher here, at 9.4\% for $N=9$ and 7.2\% for $N=14$. However, calculating the Wulff shape by computing all unique surfaces with Miller indices smaller than 1 or 2 (as commonly done) results in even higher errors of 14 and 13.2\%, and at costs of 7 and 13 calculations, respectively.

For Al$_3$Pt$_2$ (Fig.~\ref{fig:Al2Pt3_Wulff}), where the correlation is in the bottom six of our test set, low-index facets dominate the Wulff shape. Here, MAE starts off relatively high at $N=3$ with 11.2\% and as we add more surfaces, drops to 6.9\% at $N=13$ mostly due to the (100) facet appearing which also contributes to the final shape. At $N=27$, we see contributions from (111) and (2-12) which remain in the final Wulff shape, and at this value of $N$, the MAE drops to 3.6\%. Calculating up to MMI 1 or 2, however, results in MAE of 5.2 and 0.0\%, although at relatively high costs of 11 and 23 calculations, respectively. For this system, this ends up being the better approach, providing higher accuracy at fewer calculations.

For $\gamma$-TiAl (Fig.~\ref{fig:TiAl_Wulff}), where we have no correlation and do not expect this method to work well, we expect similar problems. Indeed \textsc{SurfFlow} misses the lowest energy surface, (110), which also contributes to the Wulff shape if we set $N<15$. However, the second lowest energy surface, (101) is immediately captured, as is the high energy (001) facet. The MAEs for $N=3$ and $N=9$ are decent at 9.1 and 6.8\%, and at $N=15$, we have the correct Wulff shape, even though there are 29 distinct surfaces for $\mathrm{MMI}=3$. However, computing up to MMI 1 or 2 is more advantageous here as well, with MAE of 3.2 and 2.4\%, at costs of 5 and 12 calculations respectively.

Overall, we can confidently say that utilizing the BVS is better than just relying on low-index surfaces, although low surface energy alone does not determine that a facet contributes to the Wulff shape. The average MAE (compared to the full Wulff shape for MMI 3) for all materials is 17.1\% if an MMI of 1 is used and 12.5\% if MMI is 2. Using the same number of calculations per material as for MMI 1, or 2, but using the BVS predictions to select them, the average MAE goes down to 9.0 and 2.7\%, respectively.

Fig.~\ref{fig:mae_barplot_wrt_coverage} shows a histogram of materials with respect to the coverage\footnote{In this context, coverage pertains to the ratio between the number of surfaces under consideration for a given material and the total number of surfaces available for that material.\label{footnote:coverage}} required to reach 5\% of MAE when compared to the full Wulff shape at 100\% coverage. The results show that a 40\% coverage is sufficient to reach this threshold for the vast majority of materials studied ($\sim75\%$), while only 4 of the 36 materials need more than 60\%.

Finally, the lowest-energy orientation of a crystal is also of considerable interest. In the study of 36 materials, the BVS ranking correctly predicts the surface with the lowest overall energy in 67\% of the cases, and in over 85\%, it is within the lowest 3 predictions (see inset in Fig.~\ref{fig:mae_barplot_wrt_coverage}). If finding the lowest energy surface for a set of materials is the primary concern, SurfFlow's BVS filtering is thus an excellent tool to save computation time.

\subsection{Benchmarking the \textsc{SurfFlow} workflow}
\label{sub:benchmarks}

As mentioned in section~\ref{sec:int}, it is hard to measure surface energies experimentally. Thus, it is not easy to benchmark our workflow using hard experimental data as a reference.

We have decided to first validate our workflow using some monoatomic systems with data from the materials project as reference~\cite{Tran2016}. The agreement is excellent other than for some outliers which hint at some problems in the data of reference~\cite{Tran2016} since our results match other previous studies. A detailed discussion of these results can be found in S10 of the SM.

A more rigorous test of the important capability to treat multi-component systems is to compare our calculated Wulff shapes with the experimentally determined and previously calculated nanoparticle shape of the anatase and rutile phases of TiO$_2$ (mp-390 and mp-2657). These oxide materials potentially feature polar surfaces and reconstructions, and thus present a particularly hard challenge for \textsc{SurfFlow}'s algorithms. They are also well-studied systems, both experimentally and especially theoretically (see e.g.~the reviews by Diebold and Liu et al.~\cite{Diebold2003, Liu2014}).

For these calculations we have employed the defaults of \textsc{SurfFlow} (see S5.1 in the SM), disregarded polar surfaces, and utilized the bond valence sum for pre-screening, only calculating the lowest 10 BVS terminations with DFT. 

For anatase, \textsc{SurfFlow} predicts the (101) facet to be dominant (99\%) in the Wulff shape and also having the lowest surface energy of all facets at \unit[0.41]{J/m$^2$} which is in good agreement with previous results of \unit[0.44]{J/m$^2$}~\cite{Lazzeri2002, Gong2006}.
The only other surface contributing to the Wulff shape is a small (001) facet, which has a considerably higher surface energy of \unit[0.95]{J/m$^2$}~(\unit[0.90]{J/m$^2$}~\cite{Lazzeri2002}). Nanoparticles with this shape were already grown more than 20 years ago by Penn and Banfield~\cite{Penn1999}.
Further calculated low energy surfaces are the (100) and (201) facets, at \unit[0.51]{J/m$^2$} (0.53~\cite{Lazzeri2002}) and \unit[0.63]{J/m$^2$}, respectively. Indeed, TiO$_2$ anatase nanoparticles with exposed (201) facets have been grown previously~\cite{Wu2011, Wu2013}. The other high-index surfaces showing relatively low surface energy we calculated are (112), (310), and (103). Another termination of the (201) facet with low BVS, but featuring an undercoordinated oxygen atom did not converge, so only 9 surface energies are reported in table~\ref{tab:tio2_surfen}. The correlation between the BVS and the DFT computed surface energy is very high at $r_\mathrm{anatase}=0.96$.

\begin{table}[h!]
\centering
\begin{tabular}{cccccc}
\toprule
\multicolumn{3}{c}{anatase}  & \multicolumn{3}{c}{rutile} \\
\cmidrule(r){1-3} \cmidrule(r){4-6} \\
hkl & $\gamma$ & AF & hkl & $\gamma$ & AF \\
\midrule
101 & 0.41 & 98.9 & 110 & 0.29 & 82.1 \\
100 & 0.51 & - & 100 &  0.58 & - \\
201 & 0.63 & - & 311 & 0.75 & - \\
112 & 0.71 & - & 201 & 0.79 & 8.5 \\
310 & 0.74 & - & 211 & 0.82 & - \\
210 & 0.89 & - & 331 & 0.88 & - \\
103 & 0.91 & - & 101 & 0.93 & - \\
001 & 0.95 & 1.1 & 332 & 0.97 & 9.4 \\
110 & 1.00 & - & 320 & 1.25 & - \\
\bottomrule
\end{tabular}
\caption{Surface energies $\gamma$ in [J/m$^2$] and Wulff shape area fractions (AF) in \% for anatase and rutile TiO$_2$.}
\label{tab:tio2_surfen}
\end{table}

For rutile, (110) is found to be the dominant facet, occupying 82\% of the total area of the Wulff shape, with a surface energy of \unit[0.29]{J/m$^2$}. Two more facets, (201) at \unit[0.80]{J/m$^2$}, and (101) at \unit[0.97]{J/m$^2$} each contribute about 9\% to the Wulff shape. In literature, only low index surfaces (MMI=1) are calculated, and the Wulff shapes computed feature the (110), (101), (100), and (001) facets~\cite{Ramamoorthy1994, Jiang2018}. We do calculate a low surface energy of \unit[0.58]{J/m$^2$} for the (100) facet, but it does not contribute to the Wulff shape, while the (001) surface is not calculated because its BVS sum is not within the lowest 10 surfaces.

\textsc{SurfFlow}'s result for the (110) facet is in line with a previous PBE result of \unit[0.31]{J/m$^2$}~\cite{Lazzeri2002}, while Perron et al.~report a Perdew \& Wang GGA value of \unit[0.48]{J/m$^2$}~\cite{Perron2007}, Bredow et al.~ find \unit[0.63]{J/m$^2$} for the same functional and Jiang and coworkers compute an even larger value of \unit[0.74]{J/m$^2$} with PBE~\cite{Jiang2018}.

This broad spread of values is due to well-known and large oscillations of the electronic properties of this surface with respect to the odd/even number of titanium layers~\cite{Bredow2004}, which makes the surface energy very hard to converge. Such oscillations are not found for the (100) surfaces. Our (110) result was computed for 4 titanium layers according to \textsc{SurfFlow}'s defaults, which yields a particularly low value of the surface energy, crowding out the (100) surface from our Wulff shape. High-resolution scanning tunneling microscopy images suggest that the (100) facet should however be present~\cite{Jiang2018}.

Again, one of the facets, now (320), did not converge, leaving us with 9 computed surface energies reported in table~\ref{tab:tio2_surfen}. Note that of the 9 computed surface energies, 6 have $\mathrm{MMI}>1$, and the correlation between BVS and computed $\gamma$ is decent at $r_\mathrm{rutile}=0.80$, indicating that those surfaces indeed have low surface energy compared to not calculated lower index surfaces.

In conclusion, we have developed \textsc{SurfFlow}, a Python package that efficiently computes surface energies and Wulff shapes for arbitrary crystals in a high-throughput manner. By employing symmetric considerations, optimizing slab thickness, and implementing a pre-screening approach based on the bond valence sum, we successfully reduce the computational cost associated with determining low-energy surface orientations and terminations. 
For all but very weakly bound materials we find a good to excellent correlation between the bond valence sum and the surface energy. Utilizing this correlation, we have shown that we can reliably construct more accurate Wulff shapes with fewer calculations than by just computing low-index surfaces.
This paper presents the algorithms and approximations utilized as well as the workflow architecture, web tools, and a comprehensive database that includes hundreds of surface energies for crystals with diverse chemistries and structures.
Notably, \textsc{SurfFlow} offers features such as streamlined job submission, automatic error correction, post-processing capabilities, and integration with an \textsc{optimade} compliant MongoDB database. 

\section{Methods}\label{sec:dft_settings}
We use the Vienna Ab-Initio Simulation Package (VASP, version 6.3.1,~\cite{Kresse1993, Kresse1996, Kresse1996a}) with the potpaw.54 potential set~\cite{Kresse1999}. The potential mapping is equivalent to the one defined in the MPRelaxSet of the Materials Project, except for tungsten, where we use W\_sv instead of the deprecated W\_pv. The user can easily overwrite this potential mapping while submitting workflows. Our package allows for easy changes to computational and structural parameters such as plane wave energy cutoff, k-point density, and slab thickness. We have nevertheless taken care to set default values that promise both accurate results and reasonable computation time. Those are e.g.~using \unit[400]{eV} as the plane wave cutoff, \unit[5.0]{\AA$^{-1}$} for the k-point density, and slab thicknesses of at least 8 layers or \unit[10]{\AA}, whatever is greater. All data presented here are computed with those defaults, or in some cases of older data, with higher values.
Smearing parameters change for the relaxations depending on material parameters (especially bandgap) and might be corrected during runs by custodian, but for total energies, we always include an extra static run using the tetrahedron method with Blöchl corrections. Our workflow is optimized to use the PBE or SCAN functionals, but LDA or other GGAs, as well as van der Waals corrections, can also be selected. The results calculated for this paper are using the PBE functional. More information about the choice of functional and pseudopotential set is available in section 8 of the SM.

\section*{Data availability}
All data described in this paper is available in an \textsc{optimade} compliant database accessible via a web app at: \href{https://surfflow-db.onrender.com/}{https://surfflow-db.onrender.com/}. For more advanced users that would like to connect their MongoDB apps to browse the database, we also provide the necessary \href{mongodb+srv://seread:VRFLgGmpNEd7T9iB@surfflow.asxl9nt.mongodb.net/}{URI}\footnote{\url{mongodb+srv://seread:VRFLgGmpNEd7T9iB@surfflow.asxl9nt.mongodb.net/} \label{URI}}. Additionally, we include a table with most of the results in S12 of the SM.

\section*{Code availability}
The \textsc{SurfFlow} code can be installed directly from PyPI with the same name, and the source code can be found on \href{https://github.com/fyalcin/surfflow}{GitHub}.

\section*{Acknowledgements}
This research was funded by the Austrian Science Fund (FWF) [P 32711]. The authors thank M. Reticcioli for fruitful discussions.

\section*{Author contributions}
\textbf{Firat Yalcin}: Methodology, Software, Validation, Formal analysis, Investigation, Data curation, Writing -- Original Draft, Writing -- Review \& Editing, Visualization;
\textbf{Michael Wolloch}: Conceptualization, Software, Resources, Writing - Review \& Editing, Supervision, Project administration, Funding acquisition

\section*{Competing interest}
F.Y.~declares no competing interest. M.W.~is a part-time employee of the VASP software GmbH.


\bibliographystyle{elsarticle-num}
\bibliography{main-bib}

\begin{thebibliography}{10}
\expandafter\ifx\csname url\endcsname\relax
  \def\url#1{\texttt{#1}}\fi
\expandafter\ifx\csname urlprefix\endcsname\relax\def\urlprefix{URL }\fi
\expandafter\ifx\csname href\endcsname\relax
  \def\href#1#2{#2} \def\path#1{#1}\fi

\bibitem{Wulff1901}
G.~Wulff, Zur frage der geschwindigkeit des wachstums und der aufl\"{o}sung der
  kristallformen, Zeitschrift für Kristallographie 34~(5) (1901) 449--530.

\bibitem{Somorjai1994}
G.~A. Somorjai,
  \href{https://doi.org/10.1146/annurev.pc.45.100194.003445}{{Surface
  Reconstruction and Catalysis}}, Annual Review of Physical Chemistry 45~(1)
  (1994) 721--751.
\newblock \href {https://doi.org/10.1146/annurev.pc.45.100194.003445}
  {\path{doi:10.1146/annurev.pc.45.100194.003445}}.
\newline\urlprefix\url{https://doi.org/10.1146/annurev.pc.45.100194.003445}

\bibitem{Oura2013Surface}
K.~Oura, V.~Lifshits, A.~Saranin, A.~Zotov, M.~Katayama, Surface science: An
  introduction, Springer Science \& Business Media, 2013, [Online; accessed
  2023-01-25].

\bibitem{Wynblatt1977}
P.~Wynblatt, R.~C. Ku, {Surface energy and solute strain energy effects in
  surface segregation}, Surface Science 65~(2) (1977) 511--531.
\newblock \href {https://doi.org/10.1016/0039-6028(77)90462-9}
  {\path{doi:10.1016/0039-6028(77)90462-9}}.

\bibitem{Seah1979}
M.~P. Seah, {Quantitative prediction of surface segregation}, Journal of
  Catalysis 57~(3) (1979) 450--457.
\newblock \href {https://doi.org/10.1016/0021-9517(79)90011-3}
  {\path{doi:10.1016/0021-9517(79)90011-3}}.

\bibitem{Polak2000}
M.~Polak, L.~Rubinovich,
  \href{https://www.sciencedirect.com/science/article/pii/S0167572999000102}{{The
  interplay of surface segregation and atomic order in alloys}}, Surface
  Science Reports 38~(4) (2000) 127--194.
\newblock \href {https://doi.org/https://doi.org/10.1016/S0167-5729(99)00010-2}
  {\path{doi:https://doi.org/10.1016/S0167-5729(99)00010-2}}.
\newline\urlprefix\url{https://www.sciencedirect.com/science/article/pii/S0167572999000102}

\bibitem{Hammer2000}
B.~Hammer, J.~K. N{\o}rskov, {Theoretical surface science and
  catalysis—calculations and concepts}, Advances in Catalysis 45~(C) (2000)
  71--129.
\newblock \href {https://doi.org/10.1016/S0360-0564(02)45013-4}
  {\path{doi:10.1016/S0360-0564(02)45013-4}}.

\bibitem{Stamenkovic2006}
V.~Stamenkovic, B.~S. Mun, K.~J. Mayrhofer, P.~N. Ross, N.~M. Markovic,
  J.~Rossmeisl, J.~Greeley, J.~K. N{\o}rskov, {Changing the activity of
  electrocatalysts for oxygen reduction by tuning the surface electronic
  structure}, Angewandte Chemie - International Edition 45~(18) (2006)
  2897--2901.
\newblock \href {https://doi.org/10.1002/anie.200504386}
  {\path{doi:10.1002/anie.200504386}}.

\bibitem{Norskov2009}
J.~K. N{\o}rskov, T.~Bligaard, J.~Rossmeisl, C.~H. Christensen,
  \href{https://doi.org/10.1038/nchem.121}{{Towards the computational design of
  solid catalysts}}, Nature Chemistry 1~(1) (2009) 37--46.
\newblock \href {https://doi.org/10.1038/nchem.121}
  {\path{doi:10.1038/nchem.121}}.
\newline\urlprefix\url{https://doi.org/10.1038/nchem.121}

\bibitem{Zhou2011}
Z.-Y. Zhou, N.~Tian, J.-T. Li, I.~Broadwell, S.-G. Sun,
  \href{http://dx.doi.org/10.1039/C0CS00176G}{Nanomaterials of high surface
  energy with exceptional properties in catalysis and energy storage}, Chem.
  Soc. Rev. 40 (2011) 4167--4185.
\newblock \href {https://doi.org/10.1039/C0CS00176G}
  {\path{doi:10.1039/C0CS00176G}}.
\newline\urlprefix\url{http://dx.doi.org/10.1039/C0CS00176G}

\bibitem{Seymour2023}
I.~D. Seymour, E.~Qu\'{e}rel, R.~H. Brugge, F.~M. Pesci, A.~Aguadero,
  \href{https://chemistry-europe.onlinelibrary.wiley.com/doi/abs/10.1002/cssc.202202215}{Understanding
  and engineering interfacial adhesion in solid-state batteries with metallic
  anodes}, ChemSusChem 16~(12) (2023) e202202215.
\newblock \href
  {http://arxiv.org/abs/https://chemistry-europe.onlinelibrary.wiley.com/doi/pdf/10.1002/cssc.202202215}
  {\path{arXiv:https://chemistry-europe.onlinelibrary.wiley.com/doi/pdf/10.1002/cssc.202202215}},
  \href {https://doi.org/https://doi.org/10.1002/cssc.202202215}
  {\path{doi:https://doi.org/10.1002/cssc.202202215}}.
\newline\urlprefix\url{https://chemistry-europe.onlinelibrary.wiley.com/doi/abs/10.1002/cssc.202202215}

\bibitem{Restuccia2023}
P.~Restuccia, G.~Losi, O.~Chehaimi, M.~Marsili, M.~C. Righi,
  \href{https://doi.org/10.1021/acsami.3c00662}{High-throughput
  first-principles prediction of interfacial adhesion energies in
  metal-on-metal contacts}, ACS Applied Materials \& Interfaces 15~(15) (2023)
  19624--19633, pMID: 37015021.
\newblock \href {https://doi.org/10.1021/acsami.3c00662}
  {\path{doi:10.1021/acsami.3c00662}}.
\newline\urlprefix\url{https://doi.org/10.1021/acsami.3c00662}

\bibitem{Obreimoff1930}
J.~W. Obreimoff, The splitting strength of mica, Proceedings of The Royal
  Society A: Mathematical, Physical and Engineering Sciences 127 (1930)
  290--297.

\bibitem{Gilman1960}
J.~J. Gilman, \href{https://doi.org/10.1063/1.1735524}{Direct measurements of
  the surface energies of crystals}, Journal of Applied Physics 31~(12) (1960)
  2208--2218.
\newblock \href {http://arxiv.org/abs/https://doi.org/10.1063/1.1735524}
  {\path{arXiv:https://doi.org/10.1063/1.1735524}}, \href
  {https://doi.org/10.1063/1.1735524} {\path{doi:10.1063/1.1735524}}.
\newline\urlprefix\url{https://doi.org/10.1063/1.1735524}

\bibitem{Kendall1987}
K.~Kendall, N.~McN.Alford, J.~D. Birchall,
  \href{https://doi.org/10.1038/325794a0}{A new method for measuring the
  surface energy of solids}, Nature 325~(6107) (1987) 794--796.
\newblock \href {https://doi.org/10.1038/325794a0}
  {\path{doi:10.1038/325794a0}}.
\newline\urlprefix\url{https://doi.org/10.1038/325794a0}

\bibitem{Kwok1999}
D.~Y. Kwok, A.~W. Neumann,
  \href{https://www.sciencedirect.com/science/article/pii/S0001868698000876}{{Contact
  angle measurement and contact angle interpretation}}, Advances in Colloid and
  Interface Science 81~(3) (1999) 167--249.
\newblock \href {https://doi.org/https://doi.org/10.1016/S0001-8686(98)00087-6}
  {\path{doi:https://doi.org/10.1016/S0001-8686(98)00087-6}}.
\newline\urlprefix\url{https://www.sciencedirect.com/science/article/pii/S0001868698000876}

\bibitem{Kozbial2014}
A.~Kozbial, Z.~Li, C.~Conaway, R.~McGinley, S.~Dhingra, V.~Vahdat, F.~Zhou,
  B.~D'Urso, H.~Liu, L.~Li, \href{https://doi.org/10.1021/la5018328}{{Study on
  the Surface Energy of Graphene by Contact Angle Measurements}}, Langmuir
  30~(28) (2014) 8598--8606.
\newblock \href {https://doi.org/10.1021/la5018328}
  {\path{doi:10.1021/la5018328}}.
\newline\urlprefix\url{https://doi.org/10.1021/la5018328}

\bibitem{Williams2006}
J.~A. Williams, H.~R. Le,
  \href{https://dx.doi.org/10.1088/0022-3727/39/12/R01}{Tribology and mems},
  Journal of Physics D: Applied Physics 39~(12) (2006) R201.
\newblock \href {https://doi.org/10.1088/0022-3727/39/12/R01}
  {\path{doi:10.1088/0022-3727/39/12/R01}}.
\newline\urlprefix\url{https://dx.doi.org/10.1088/0022-3727/39/12/R01}

\bibitem{Xiao2020}
C.~Xiao, B.-A. Lu, P.~Xue, N.~Tian, Z.-Y. Zhou, X.~Lin, W.-F. Lin, S.-G. Sun,
  \href{https://doi.org/10.1016/j.joule.2020.10.002}{High-index-facet- and
  high-surface-energy nanocrystals of metals and metal oxides as highly
  efficient catalysts}, Joule 4~(12) (2020) 2562--2598.
\newblock \href {https://doi.org/10.1016/j.joule.2020.10.002}
  {\path{doi:10.1016/j.joule.2020.10.002}}.
\newline\urlprefix\url{https://doi.org/10.1016/j.joule.2020.10.002}

\bibitem{Tran2016}
R.~Tran, Z.~Xu, B.~Radhakrishnan, D.~Winston, W.~Sun, K.~A. Persson, S.~P. Ong,
  \href{https://doi.org/10.1038/sdata.2016.80}{{Surface energies of elemental
  crystals}}, Scientific Data 3~(1) (2016) 160080.
\newblock \href {https://doi.org/10.1038/sdata.2016.80}
  {\path{doi:10.1038/sdata.2016.80}}.
\newline\urlprefix\url{https://doi.org/10.1038/sdata.2016.80}

\bibitem{Yang2020}
S.~Yang, I.~Bier, W.~Wen, J.~Zhan, S.~Moayedpour, N.~Marom,
  \href{https://doi.org/10.1063/5.0010615}{{Ogre: A Python package for
  molecular crystal surface generation with applications to surface energy and
  crystal habit prediction}}, Journal of Chemical Physics 152~(24) (2020).
\newblock \href {https://doi.org/10.1063/5.0010615}
  {\path{doi:10.1063/5.0010615}}.
\newline\urlprefix\url{https://doi.org/10.1063/5.0010615}

\bibitem{Brlec2021}
K.~Brlec, D.~Davies, D.~Scanlon, {Surfaxe: Systematic surface calculations},
  Journal of Open Source Software 6~(61) (2021) 3171.
\newblock \href {https://doi.org/10.21105/joss.03171}
  {\path{doi:10.21105/joss.03171}}.

\bibitem{Palizhati2019}
A.~Palizhati, W.~Zhong, K.~Tran, Z.~W. Ulissi, {Predicting intermetallic
  surface energies with high-throughput dft and convolutional neural networks},
  ChemRxiv (2019) 1--18\href {https://doi.org/10.26434/chemrxiv.8709566}
  {\path{doi:10.26434/chemrxiv.8709566}}.

\bibitem{Moayedpour2023}
S.~Moayedpour, I.~Bier, W.~Wen, D.~Dardzinski, O.~Isayev, N.~Marom,
  \href{https://doi.org/10.1021/acs.jpcc.3c02384}{Structure prediction of
  epitaxial organic interfaces with ogre, demonstrated for
  tetracyanoquinodimethane (tcnq) on tetrathiafulvalene (ttf)}, The Journal of
  Physical Chemistry C 127~(21) (2023) 10398--10410.
\newblock \href {https://doi.org/10.1021/acs.jpcc.3c02384}
  {\path{doi:10.1021/acs.jpcc.3c02384}}.
\newline\urlprefix\url{https://doi.org/10.1021/acs.jpcc.3c02384}

\bibitem{ONG2013314}
S.~P. Ong, W.~D. Richards, A.~Jain, G.~Hautier, M.~Kocher, S.~Cholia,
  D.~Gunter, V.~L. Chevrier, K.~A. Persson, G.~Ceder,
  \href{https://www.sciencedirect.com/science/article/pii/S0927025612006295}{{Python
  Materials Genomics (pymatgen): A robust, open-source python library for
  materials analysis}}, Computational Materials Science 68 (2013) 314--319.
\newblock \href {https://doi.org/10.1016/j.commatsci.2012.10.028}
  {\path{doi:10.1016/j.commatsci.2012.10.028}}.
\newline\urlprefix\url{https://www.sciencedirect.com/science/article/pii/S0927025612006295}

\bibitem{MATHEW2017140}
K.~Mathew, J.~H. Montoya, A.~Faghaninia, S.~Dwarakanath, M.~Aykol, H.~Tang,
  I.~heng Chu, T.~Smidt, B.~Bocklund, M.~Horton, J.~Dagdelen, B.~Wood, Z.~K.
  Liu, J.~Neaton, S.~P. Ong, K.~Persson, A.~Jain,
  \href{https://www.sciencedirect.com/science/article/pii/S0927025617303919}{{Atomate:
  A high-level interface to generate, execute, and analyze computational
  materials science workflows}}, Computational Materials Science 139 (2017)
  140--152.
\newblock \href {https://doi.org/10.1016/j.commatsci.2017.07.030}
  {\path{doi:10.1016/j.commatsci.2017.07.030}}.
\newline\urlprefix\url{https://www.sciencedirect.com/science/article/pii/S0927025617303919}

\bibitem{jain:15}
A.~Jain, S.~P. Ong, W.~Chen, B.~Medasani, X.~Qu, M.~Kocher, M.~Brafman,
  G.~Petretto, G.-M. Rignanese, G.~Hautier, D.~Gunter, K.~A. Persson,
  \href{https://onlinelibrary.wiley.com/doi/abs/10.1002/cpe.3505}{{FireWorks: a
  dynamic workflow system designed for high-throughput applications}},
  Concurrency and Computation: Practice and Experience 27~(17) (2015)
  5037--5059.
\newblock \href {https://doi.org/https://doi.org/10.1002/cpe.3505}
  {\path{doi:https://doi.org/10.1002/cpe.3505}}.
\newline\urlprefix\url{https://onlinelibrary.wiley.com/doi/abs/10.1002/cpe.3505}

\bibitem{Kresse1993}
G.~Kresse, J.~Hafner,
  \href{https://link.aps.org/doi/10.1103/PhysRevB.47.558}{Ab initio molecular
  dynamics for liquid metals}, Phys. Rev. B 47 (1993) 558--561.
\newblock \href {https://doi.org/10.1103/PhysRevB.47.558}
  {\path{doi:10.1103/PhysRevB.47.558}}.
\newline\urlprefix\url{https://link.aps.org/doi/10.1103/PhysRevB.47.558}

\bibitem{Kresse1996}
G.~Kresse, J.~Furthm{\"{u}}ller, {Efficiency of ab-initio total energy
  calculations for metals and semiconductors using a plane-wave basis set},
  Computational Materials Science 6~(1) (1996) 15--50.
\newblock \href {https://doi.org/10.1016/0927-0256(96)00008-0}
  {\path{doi:10.1016/0927-0256(96)00008-0}}.

\bibitem{Kresse1996a}
G.~Kresse, J.~Furthm{\"{u}}ller, {Efficient iterative schemes for ab initio
  total-energy calculations using a plane-wave basis set}, Physical Review B -
  Condensed Matter and Materials Physics 54~(16) (1996) 11169--11186.
\newblock \href {https://doi.org/10.1103/PhysRevB.54.11169}
  {\path{doi:10.1103/PhysRevB.54.11169}}.

\bibitem{SUN201353}
W.~Sun, G.~Ceder,
  \href{https://www.sciencedirect.com/science/article/pii/S003960281300160X}{{Efficient
  creation and convergence of surface slabs}}, Surface Science 617 (2013)
  53--59.
\newblock \href {https://doi.org/10.1016/j.susc.2013.05.016}
  {\path{doi:10.1016/j.susc.2013.05.016}}.
\newline\urlprefix\url{https://www.sciencedirect.com/science/article/pii/S003960281300160X}

\bibitem{Jia2019}
T.~Jia, Z.~Zeng, H.~Paudel, D.~J. Senor, Y.~Duan, {First-principles study of
  the surface properties of $\gamma$-LiAlO2: Stability and tritium adsorption},
  Journal of Nuclear Materials 522 (2019) 1--10.
\newblock \href {https://doi.org/10.1016/j.jnucmat.2019.05.007}
  {\path{doi:10.1016/j.jnucmat.2019.05.007}}.

\bibitem{Jia2021}
T.~Jia, D.~J. Senor, Y.~Duan, First-principles study of the surface properties
  of {{LiAl5O8}}: {{Stability}} and tritiated water formation, Journal of
  Nuclear Materials 555 (2021) 153111.
\newblock \href {https://doi.org/10.1016/j.jnucmat.2021.153111}
  {\path{doi:10.1016/j.jnucmat.2021.153111}}.

\bibitem{Zhang2016}
J.~Zhang, Y.~Zhang, K.~Tse, B.~Deng, H.~Xu, J.~Zhu, {New approaches for
  calculating absolute surface energies of wurtzite (0001)/(000 1): A study of
  ZnO and GaN}, Journal of Applied Physics 119~(20) (2016) 0--8.
\newblock \href {https://doi.org/10.1063/1.4952395}
  {\path{doi:10.1063/1.4952395}}.

\bibitem{Ma2019}
C.~Ma, W.~Jin, X.~Duan, X.~Ma, H.~Han, Z.~Zhang, J.~Yu, Y.~Wu,
  \href{https://doi.org/10.1016/j.jallcom.2018.09.194}{{From the absolute
  surface energy to the stabilization mechanism of high index polar surface in
  wurtzite structure: The case of ZnO}}, Journal of Alloys and Compounds 772
  (2019) 482--488.
\newblock \href {https://doi.org/10.1016/j.jallcom.2018.09.194}
  {\path{doi:10.1016/j.jallcom.2018.09.194}}.
\newline\urlprefix\url{https://doi.org/10.1016/j.jallcom.2018.09.194}

\bibitem{Chetty1992}
N.~Chetty, R.~M. Martin, {First-principles energy density and its applications
  to selected polar surfaces}, Physical Review B 45~(11) (1992) 6074--6088.
\newblock \href {https://doi.org/10.1103/PhysRevB.45.6074}
  {\path{doi:10.1103/PhysRevB.45.6074}}.

\bibitem{Kaminski2017}
J.~W. Kaminski, P.~Kratzer, C.~Ratsch, {Towards a standardized setup for
  surface energy calculations}, Physical Review B 95~(8) (2017) 1--11.
\newblock \href {https://doi.org/10.1103/PhysRevB.95.085408}
  {\path{doi:10.1103/PhysRevB.95.085408}}.

\bibitem{Bruno2021}
M.~Bruno, S.~Ghignone, {A new computational strategy to calculate the surface
  energy of a dipolar crystal surface}, CrystEngComm 23~(27) (2021) 4791--4798.
\newblock \href {https://doi.org/10.1039/d1ce00403d}
  {\path{doi:10.1039/d1ce00403d}}.

\bibitem{Heifets2001}
E.~Heifets, R.~I. Eglitis, E.~A. Kotomin, J.~Maier, G.~Borstel, {\emph{Ab
  Initio}} modeling of surface structure for {{SrTiO}} 3 perovskite crystals,
  Physical Review B 64~(23) (2001) 235417.
\newblock \href {https://doi.org/10.1103/PhysRevB.64.235417}
  {\path{doi:10.1103/PhysRevB.64.235417}}.

\bibitem{Tian2018}
X.~Tian, T.~Wang, L.~Fan, Y.~Wang, H.~Lu, Y.~Mu,
  \href{http://dx.doi.org/10.1016/j.apsusc.2017.08.172
  https://doi.org/10.1016/j.apsusc.2017.08.172}{{A DFT based method for
  calculating the surface energies of asymmetric MoP facets}}, Applied Surface
  Science 427 (2018) 357--362.
\newblock \href {https://doi.org/10.1016/j.apsusc.2017.08.172}
  {\path{doi:10.1016/j.apsusc.2017.08.172}}.
\newline\urlprefix\url{http://dx.doi.org/10.1016/j.apsusc.2017.08.172
  https://doi.org/10.1016/j.apsusc.2017.08.172}

\bibitem{Eglitis2007}
R.~I. Eglitis, D.~Vanderbilt, {\emph{Ab Initio}} calculations of {{Ba Ti O}} 3
  and {{Pb Ti O}} 3 (001) and (011) surface structures, Physical Review B
  76~(15) (2007) 155439.
\newblock \href {https://doi.org/10.1103/PhysRevB.76.155439}
  {\path{doi:10.1103/PhysRevB.76.155439}}.

\bibitem{Noguera2000}
C.~Noguera, {Polar oxide surfaces}, Journal of Physics Condensed Matter 12~(31)
  (2000).
\newblock \href {https://doi.org/10.1088/0953-8984/12/31/201}
  {\path{doi:10.1088/0953-8984/12/31/201}}.

\bibitem{Goniakowski2008}
J.~Goniakowski, F.~Finocchi, C.~Noguera, {Polarity of oxide surfaces and
  nanostructures}, Reports on Progress in Physics 71~(1) (2008).
\newblock \href {https://doi.org/10.1088/0034-4885/71/1/016501}
  {\path{doi:10.1088/0034-4885/71/1/016501}}.

\bibitem{Dreyer2014}
C.~E. Dreyer, A.~Janotti, C.~G. {Van De Walle}, {Absolute surface energies of
  polar and nonpolar planes of GaN}, Physical Review B - Condensed Matter and
  Materials Physics 89~(8) (2014) 1--4.
\newblock \href {https://doi.org/10.1103/PhysRevB.89.081305}
  {\path{doi:10.1103/PhysRevB.89.081305}}.

\bibitem{Methfessel1992}
M.~Methfessel, D.~Hennig, M.~Scheffler, {Calculated surface energies of the 4 d
  transition metals: A study of bond-cutting models}, Applied Physics A Solids
  and Surfaces 55~(5) (1992) 442--448.
\newblock \href {https://doi.org/10.1007/BF00348331}
  {\path{doi:10.1007/BF00348331}}.

\bibitem{Galanakis2002}
I.~Galanakis, N.~Papanikolaou, P.~H. Dederichs, {Applicability of the
  broken-bond rule to the surface energy of the fcc metals}, Surface Science
  511~(1-3) (2002) 1--12.
\newblock \href {http://arxiv.org/abs/0110236} {\path{arXiv:0110236}}, \href
  {https://doi.org/10.1016/S0039-6028(02)01547-9}
  {\path{doi:10.1016/S0039-6028(02)01547-9}}.

\bibitem{Gao2014}
Z.~Y. Gao, W.~Sun, Y.~H. Hu,
  \href{http://dx.doi.org/10.1016/S1003-6326(14)63428-2}{{Mineral cleavage
  nature and surface energy: Anisotropic surface broken bonds consideration}},
  Transactions of Nonferrous Metals Society of China (English Edition) 24~(9)
  (2014) 2930--2937.
\newblock \href {https://doi.org/10.1016/S1003-6326(14)63428-2}
  {\path{doi:10.1016/S1003-6326(14)63428-2}}.
\newline\urlprefix\url{http://dx.doi.org/10.1016/S1003-6326(14)63428-2}

\bibitem{Etxebarria2005}
I.~Etxebarria, J.~M. Perez-Mato, A.~Garc{\'{i}}a, P.~Blaha, K.~Schwarz,
  J.~Rodriguez-Carvajal, {Comparison of empirical bond-valence and
  first-principles energy calculations for a complex structural instability},
  Physical Review B - Condensed Matter and Materials Physics 72~(17) (2005)
  1--8.
\newblock \href {https://doi.org/10.1103/PhysRevB.72.174108}
  {\path{doi:10.1103/PhysRevB.72.174108}}.

\bibitem{Brown2009}
I.~D. Brown, {Recent developments in the methods and applications of the bond
  valence model}, Chemical Reviews 109~(12) (2009) 6858--6919.
\newblock \href {https://doi.org/10.1021/cr900053k}
  {\path{doi:10.1021/cr900053k}}.

\bibitem{Ma2020}
H.~Ma, Y.~Jiao, W.~Guo, X.~Liu, Y.~Li, X.-D. Wen,
  \href{https://doi.org/10.1021/acs.jpcc.0c03537}{Predicting crystal morphology
  using a geometric descriptor: A comparative study of elemental crystals with
  high-throughput dft calculations}, The Journal of Physical Chemistry C
  124~(29) (2020) 15920--15927.
\newblock \href {https://doi.org/10.1021/acs.jpcc.0c03537}
  {\path{doi:10.1021/acs.jpcc.0c03537}}.
\newline\urlprefix\url{https://doi.org/10.1021/acs.jpcc.0c03537}

\bibitem{Sang:2012}
X.~Sang, A.~Kulovits, G.~Wang, J.~Wiezorek,
  \href{https://doi.org/10.1080/14786435.2012.709324}{High precision electronic
  charge density determination for l10-ordered $\gamma$-tial by quantitative
  convergent beam electron diffraction}, Philosophical Magazine 92~(35) (2012)
  4408--4424.
\newblock \href {https://doi.org/10.1080/14786435.2012.709324}
  {\path{doi:10.1080/14786435.2012.709324}}.
\newline\urlprefix\url{https://doi.org/10.1080/14786435.2012.709324}

\bibitem{Diebold2003}
U.~Diebold,
  \href{https://www.sciencedirect.com/science/article/pii/S0167572902001000}{The
  surface science of titanium dioxide}, Surface Science Reports 48~(5) (2003)
  53--229.
\newblock \href {https://doi.org/https://doi.org/10.1016/S0167-5729(02)00100-0}
  {\path{doi:https://doi.org/10.1016/S0167-5729(02)00100-0}}.
\newline\urlprefix\url{https://www.sciencedirect.com/science/article/pii/S0167572902001000}

\bibitem{Liu2014}
G.~Liu, H.~G. Yang, J.~Pan, Y.~Q. Yang, G.~Q.~M. Lu, H.-M. Cheng,
  \href{https://doi.org/10.1021/cr400621z}{Titanium dioxide crystals with
  tailored facets}, Chemical Reviews 114~(19) (2014) 9559--9612.
\newblock \href {https://doi.org/10.1021/cr400621z}
  {\path{doi:10.1021/cr400621z}}.
\newline\urlprefix\url{https://doi.org/10.1021/cr400621z}

\bibitem{Lazzeri2002}
M.~Lazzeri, A.~Vittadini, A.~Selloni,
  \href{https://link.aps.org/doi/10.1103/PhysRevB.65.119901}{Erratum: Structure
  and energetics of stoichiometric ${\mathrm{tio}}_{2}$ anatase surfaces [phys.
  rev. b 63, 155409 (2001)]}, Phys. Rev. B 65 (2002) 119901.
\newblock \href {https://doi.org/10.1103/PhysRevB.65.119901}
  {\path{doi:10.1103/PhysRevB.65.119901}}.
\newline\urlprefix\url{https://link.aps.org/doi/10.1103/PhysRevB.65.119901}

\bibitem{Gong2006}
X.-Q. Gong, A.~Selloni, M.~Batzill, U.~Diebold,
  \href{https://doi.org/10.1038/nmat1695}{Steps on anatase tio2(101)}, Nature
  Materials 5~(8) (2006) 665--670.
\newblock \href {https://doi.org/10.1038/nmat1695}
  {\path{doi:10.1038/nmat1695}}.
\newline\urlprefix\url{https://doi.org/10.1038/nmat1695}

\bibitem{Penn1999}
R.~Penn, J.~F. Banfield,
  \href{https://www.sciencedirect.com/science/article/pii/S001670379900037X}{Morphology
  development and crystal growth in nanocrystalline aggregates under
  hydrothermal conditions: insights from titania}, Geochimica et Cosmochimica
  Acta 63~(10) (1999) 1549--1557.
\newblock \href {https://doi.org/https://doi.org/10.1016/S0016-7037(99)00037-X}
  {\path{doi:https://doi.org/10.1016/S0016-7037(99)00037-X}}.
\newline\urlprefix\url{https://www.sciencedirect.com/science/article/pii/S001670379900037X}

\bibitem{Wu2011}
H.~B. Wu, J.~S. Chen, X.~W.~D. Lou, H.~H. Hng,
  \href{http://dx.doi.org/10.1039/C1NR10854A}{Asymmetric anatase tio2
  nanocrystals with exposed high-index facets and their excellent lithium
  storage properties}, Nanoscale 3 (2011) 4082--4084.
\newblock \href {https://doi.org/10.1039/C1NR10854A}
  {\path{doi:10.1039/C1NR10854A}}.
\newline\urlprefix\url{http://dx.doi.org/10.1039/C1NR10854A}

\bibitem{Wu2013}
L.~Wu, H.~B. Jiang, F.~Tian, Z.~Chen, C.~Sun, H.~G. Yang,
  \href{http://dx.doi.org/10.1039/C3CC38105F}{Ti0.89si0.11o2 single crystals
  bound by high-index {201} facets showing enhanced visible-light
  photocatalytic hydrogen evolution}, Chem. Commun. 49 (2013) 2016--2018.
\newblock \href {https://doi.org/10.1039/C3CC38105F}
  {\path{doi:10.1039/C3CC38105F}}.
\newline\urlprefix\url{http://dx.doi.org/10.1039/C3CC38105F}

\bibitem{Ramamoorthy1994}
M.~Ramamoorthy, D.~Vanderbilt, R.~D. King-Smith,
  \href{https://link.aps.org/doi/10.1103/PhysRevB.49.16721}{First-principles
  calculations of the energetics of stoichiometric ${\mathrm{tio}}_{2}$
  surfaces}, Phys. Rev. B 49 (1994) 16721--16727.
\newblock \href {https://doi.org/10.1103/PhysRevB.49.16721}
  {\path{doi:10.1103/PhysRevB.49.16721}}.
\newline\urlprefix\url{https://link.aps.org/doi/10.1103/PhysRevB.49.16721}

\bibitem{Jiang2018}
F.~Jiang, L.~Yang, D.~Zhou, G.~He, J.~Zhou, F.~Wang, Z.-G. Chen,
  \href{https://www.sciencedirect.com/science/article/pii/S0169433217336371}{First-principles
  atomistic wulff constructions for an equilibrium rutile tio2 shape modeling},
  Applied Surface Science 436 (2018) 989--994.
\newblock \href {https://doi.org/https://doi.org/10.1016/j.apsusc.2017.12.050}
  {\path{doi:https://doi.org/10.1016/j.apsusc.2017.12.050}}.
\newline\urlprefix\url{https://www.sciencedirect.com/science/article/pii/S0169433217336371}

\bibitem{Perron2007}
H.~Perron, C.~Domain, J.~Roques, R.~Drot, E.~Simoni, H.~Catalette,
  \href{https://doi.org/10.1007/s00214-006-0189-y}{Optimisation of accurate
  rutile tio2 (110), (100), (101) and (001) surface models from periodic dft
  calculations}, Theoretical Chemistry Accounts 117~(4) (2007) 565--574.
\newblock \href {https://doi.org/10.1007/s00214-006-0189-y}
  {\path{doi:10.1007/s00214-006-0189-y}}.
\newline\urlprefix\url{https://doi.org/10.1007/s00214-006-0189-y}

\bibitem{Bredow2004}
T.~Bredow, L.~Giordano, F.~Cinquini, G.~Pacchioni,
  \href{https://link.aps.org/doi/10.1103/PhysRevB.70.035419}{Electronic
  properties of rutile $\mathrm{Ti}{\mathrm{o}}_{2}$ ultrathin films: Odd-even
  oscillations with the number of layers}, Phys. Rev. B 70 (2004) 035419.
\newblock \href {https://doi.org/10.1103/PhysRevB.70.035419}
  {\path{doi:10.1103/PhysRevB.70.035419}}.
\newline\urlprefix\url{https://link.aps.org/doi/10.1103/PhysRevB.70.035419}

\bibitem{Kresse1999}
G.~Kresse, D.~Joubert,
  \href{https://link.aps.org/doi/10.1103/PhysRevB.59.1758}{From ultrasoft
  pseudopotentials to the projector augmented-wave method}, Phys. Rev. B 59
  (1999) 1758--1775.
\newblock \href {https://doi.org/10.1103/PhysRevB.59.1758}
  {\path{doi:10.1103/PhysRevB.59.1758}}.
\newline\urlprefix\url{https://link.aps.org/doi/10.1103/PhysRevB.59.1758}

\end{thebibliography}

\end{document}